\documentclass[aps,prl,twocolumn,showpacs,longbibliography]{revtex4-1}

\usepackage{graphicx}
\usepackage[colorlinks,linkcolor=red,citecolor=blue]{hyperref}
\usepackage{breakurl}
\usepackage[ansinew]{inputenc}
\usepackage{array}
\usepackage{color}
\usepackage{amsmath}
\usepackage{amsxtra}
\usepackage{amstext}
\usepackage{amssymb}
\usepackage{physics}
\usepackage{latexsym}
\usepackage{float}
\newcommand {\UA}[1]{ {\textcolor{blue}{#1}}}

\begin{document} 

\title{Raman interferometry between autoionizing states to probe ultrafast wavepacket dynamics with high spectral resolution}

\author{A. Plunkett$^1$, M. A. Alarc\'on$^2$, J. K. Wood$^3$, C. H. Greene$^{2}$, A. Sandhu$^{1,3}$}
\affiliation{$^1$Department of Physics, University of Arizona, Tucson, Arizona 85721, USA}
\affiliation{$^2$Department of Physics and Astronomy, Purdue University, West Lafayette, Indiana 47907, USA}
\affiliation{$^3$College of Optical Sciences, University of Arizona, Tucson, Arizona 85721, USA}
\date{\today}
\email{asandhu@arizona.edu}

\begin{abstract}
Photoelectron interferometry with femto- and atto-second light pulses is a powerful probe of the fast electron wavepacket dynamics, albeit it has practical limitations on the energy resolution. We show that one can simultaneously obtain both high temporal and spectral resolution by stimulating Raman interferences with one light pulse and monitoring the differential changes in the electron yield in a separate step. Applying this spectroscopic approach to the autoionizing states of argon, we experimentally resolved its electronic composition and time-evolution in exquisite detail. Theoretical calculations show remarkable agreement with the observations and shed light on the  light-matter interaction parameters. Using appropriate Raman probing and delayed detection steps, this technique enables highly differential probing and control of electron dynamics in complex systems. 
\end{abstract}

\maketitle

%% Introduction
Photoelectron wavepacket interferometry has been extensively applied to probe the electronic properties of atoms and molecules, to characterize the amplitudes and phases of constituent wavefunctions~\cite{Remetter2006, Klunder2013, Liu2012, Han2018,Forbes2018,Noordam1997}, and probe the strong field dynamics~\cite{Li2019, Villeneuve2017}, autoionization~\cite{Busto2018, Gruson2016,Ramswell1998}, correlated electron dynamics~\cite{Mansson2014, Usenko2020, Feist2011}, and molecular dynamics~\cite{Gonzalez-Castrillo2020}. Typically, a bound atomic or molecular wavepacket can be prepared by exciting the system with a broadband femtosecond or attosecond pulse, followed by a delayed ionizing probe light pulse which interferes various components of the wavepacket in the continuum. The interference pattern can be resolved in kinetic energy, and its evolution exhibits quantum beating in time delay due the phase differences between the states. Beat frequencies obtained from the Fourier analysis of the signal can then be used to deduce information about the composition of the wavepacket. However, the beat frequency resolution is constrained by the practical limitations on the time-delay range that can be explored in a typical experiment. On the other hand, electron kinetic energy resolution is constrained by the probe pulse duration. These issues limit the amount of information that can be extracted from spectrograms, especially when dealing with a multitude of states with unknown energies and potentially overlapping beat frequencies. 
%The interpretation of such data sets has often required a separate high-resolution study of the photoabsorption spectra. 

\begin{figure}
\includegraphics[width=8cm]{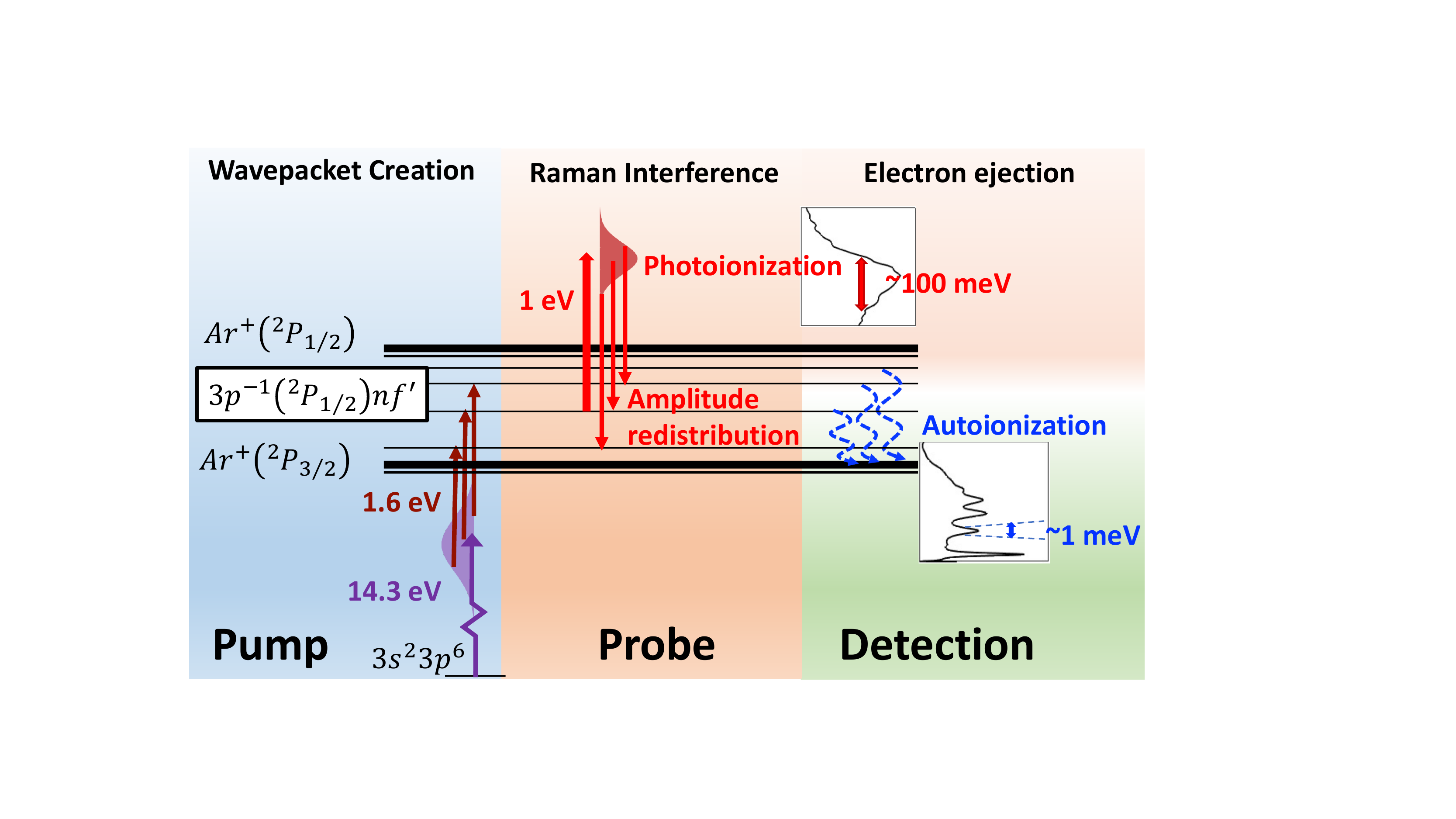}
\caption{Two color (EUV+NIR) `pump' pulse excites an $nf'$ autoionizing wavepacket in the argon atom. A time-delayed femtosecond SWIR `probe' pulse induces impulsive Raman transitions between the constituents of the wavepacket, redistributing the population within the wavepacket. Electronic states undergo slow autoionization producing sharp features in the electron spectra. Analysis of differential changes in this delayed signal as function of pump-probe delay provides both time and energy resolution.}
\label{fig:Setup}
\end{figure}

We introduce an alternate approach, that employs impulsive stimulated Raman interferences, followed by a deferred detection, to investigate the electron dynamics with both high temporal \emph{and} spectral resolution. After an excited wavepacket is prepared, a time-delayed Raman probe pulse redistributes the wavepacket amplitudes, setting up a time-dependent quantum beat in the bound or metastable state manifold, allowing femtosecond-resolved probing and control of the electron dynamics. For the actual electron signal acquisition, we rely on a deferred detection through the natural decay of the electronic states or by using a separate narrow-band ionization pulse. This long-timescale detection provides the high energy resolution, enabling the direct identification of the electronic constituents of the wavepacket. Furthermore, we achieve high sensitivity by analyzing the differential changes in the electron signal, i.e., by extracting the Raman probe induced modification of the total signal. Combination of these advances can enable new insights into the electronic structure and quantum dynamics in atoms and molecules, while also elucidating the nature of light-induced couplings in exquisite detail. 

To demonstrate the power of this technique we apply it to investigate a wavepacket composed of the autoionizing states of argon, as illustrated in Fig.~\ref{fig:Setup}. Argon ion features a spin-orbit splitting of 180 meV, with the $(^2P_{3/2})$ ground state and $(^2P_{1/2})$ excited state lying 15.76~eV and 15.94~eV above the neutral ground state, respectively \citep{NIST}. Between these two thresholds, electrons exist autoionizing resonances where the Rydberg electron is attached to an ``ion-core" state $^2P_{1/2}$. A two-color pump consisting of a few-fs extreme ultraviolet (EUV) emission centered at 14.3~eV and a 40~fs near-infrared (NIR) pulse 1.6~eV was used to coherently launch an `optically-dark' autoionizing $nf'$ wavepacket within this argon spin-orbit-excited manifold. A time-delayed $\sim$60~fs (FWHM), short-wave-infrared (SWIR) probe is used induce Raman transitions between the constituents of the wavepacket. Due to multi-electron interactions, the states relax via autoionization, simultaneously releasing the Rydberg electron and rearranging the core state to the lower-energy spin-orbit configuration: $3s^2 3p^5(^2P_{1/2})nl \rightarrow 3s^2 3p^5(^2P_{3/2})+e^-$. This natural autoionization decay, which produces sharp electron peaks between 40-150~meV, is used as detection channel. The population redistribution by the Raman probe causes differential changes in the electron signal, which can be detected by subtracting the spectra obtained with and without SWIR Raman probe. Fourier analysis of the energy resolved differential signals allows us to unambiguously determine the constituent states of the wavepacket even when the beat frequencies are overlapping.

To obtain the light pulses, we employed a Ti:Sapphire 780 nm NIR laser amplifier with 2~mJ pulse energy, 40~fs pulse duration. Half of the NIR pulse energy drives the EUV generation in xenon filled gas cell. EUV pulse was focused onto an effusive argon jet with a toroidal mirror, along with the co-propagating NIR driver. The other half of the NIR pulse from the amplifier was used to produce the SWIR pulse with an optical parametric amplifier. SWIR pulse was focused onto the argon target with a 50~cm lens. The pump (EUV+NIR) and time-delayed probe (SWIR) were collinearly recombined on a mirror with a hole before the target chamber. The photoelectrons and autoionization electrons were acquired using a velocity map imaging spectrometer. 

Tools from multichannel quantum defect theory (MQDT) and second order time dependent perturbation theory are used to model the light-matter interaction. Assuming that only the outer electron interacts with the field, the two-color (pump) excitation proceeds through an intermediate $3d'$ state viewed as purely attached to the $3s^23p^5 (^2P_{1/2})$ ionic core. This restricts the space of possible states that form the wave packet to bound states with total $J=0$ or $J=2$, i.e., an ionic core in state $3s^23p^5 (^2P_{1/2})$ and an outer electron with orbital angular momentum equal to $\ell=1$ or $\ell=3$. Using the MQDT parameters from Pellarin \emph{et al.} \cite{Pellarin1988} we determined the width and position of the autoionizing resonances with total $J=2$ for both $p$ and $f$ states \cite{Aymar1996, Seaton1983}. Because the $f$ states dominate by approximately two orders of magnitude, the $p$ autoionizing Rydberg states are neglected.
%For the total $J=0$ $p$ states the MQDT parameters were fitted to match the data for known bound states with this symmetry. Using equation (7) of \cite{Robicheaux1996} and using the outgoing wave boundary condition as prescribed in \cite{Aymar1996}, we determined the excitation amplitudes of the wave packet for all possible states. We found that the excitation probability to $f$ states is about two orders of magnitude larger than the $p$ states, in accordance with the propensity rule. Therefore we are justified to approximate the wave packet to only contain $f$ states.

As in \citep{Robicheaux1996} the wavepacket is described by an integral over energy, and a sum over the open channels and allowed symmetries. Since the excitation probability is highly concentrated around the narrow $f$ resonances, an approximate wavepacket is obtained by summing over the countable resonances 
\begin{equation}
    \left|\Psi_{o}\right\rangle=\mathcal{A} \sum_{n_{o}} A_{n_{o}}\left|\psi_{n_{o}}\right\rangle=\mathcal{A} \sum_{n_{o}} A_{n_{o}} \frac{u_{n_{o}, 3}(r)}{r} \Phi^{J=2}_{1/2},
\end{equation}
where $\Phi^{J=2}_{j_c}=\chi_{j_c} \ket{\{[(\ell_c,s_c)j_c,s]J_cs,\ell\}J,M}$ is the $Jcs$ coupled angular momentum state\cite{Lindsay1992}, with an ionic core of angular momentum $j_c$ and total angular momentum $J$. %Chosen by convenience as done in \cite{Lindsay1992}.
$\mathcal{A}$ is the antisymmetrization operator.
%, that has no effect since the electrons are in different regions of space, and $A_{n_o}$ are the initial amplitudes of the wave packet.

The radial part of the wave function $u_{n_o,3}(r)$ is approximated here by a hydrogenic orbital rather than by a Whittaker function, since $nf$-state quantum defects are small. %will, in general, be given by a Whittaker function evaluated at the energy of the resonance \cite{Aymar1996}. In this case, given the small quantum defect of the $nf'$ resonances, we can use the simpler Hydrogenic radial orbitals without loosing much precision insofar as the main difference between the Whittaker and Hydrogenic wave functions lies at small radii which are suppressed when calculating dipole matrix elements in the length gauge. 
Our treatment considers only the $J=2$ functions with $j_c=1/2$ and $\ell=3$, giving $J_{cs}=1$. The dependence of $A_{no}$ on $n_o$ reflects the characteristics of the driver laser, with an assumed  Gaussian spectral profile. 
%The wave packet is normalized to unity, $\braket{\psi_o}=1$ since it describes the bound part of the wave function only.

The probability amplitude for each autoionizing state in the wavepacket upon interaction with the delayed SWIR pulse is based on the perturbative expansion of the transition operator, as in Chapter 2 of \cite{Faisal1987}. The first non-vanishing contribution comes from the second order term, corresponding to a two photon Raman process.

For a laser pulse with field strength $\mathcal{E}_o$, central frequency $\omega$, polarization $\hat{\epsilon}$, time delay $t_o$ and temporal FWHM $2\sigma \ln{2}$, the time dependent perturbation is: 
\begin{equation}
 V(t)=\mathcal{E}_{o} e^{-\left((t-t_{o})/\sigma\right)^{2}} \cos (\omega t) \hat{\epsilon} \cdot \vec{r},
\end{equation}
in the length gauge. With this perturbation, the second order transition matrix element between the initial wave packet and one of the Rydberg states has the form:
%  V(t)=\mathcal{E}_{o} \exp \left\{-\left(\frac{t-t_{o}}{\sigma}\right)^{2}\right\} \cos (\omega t) \hat{\epsilon} \cdot \vec{r},
\begin{small}
\begin{equation}
    \bra{\psi_f} T^{(2)} \ket{\Psi_o} = - \mathcal{E}_o^2 \sum_{n_o} T_{f,n_o} e^{-i(E_f-E_{n_o})t_o -\frac{\Gamma_f+\Gamma_{n_o}}{2} t_o},
\end{equation}
\end{small}
where $E_i$ is the energy of the resonance and $\Gamma_i$ is the autoionization width of the state.

The amplitude $T^{(2)}_{f,n_o}$ involves a sum/integral over intermediate states of the dipole matrix elements multiplied by the spectral profile of the laser evaluated at the photon energies required for the Raman transition. The sum of zeroth- and second-order transition matrix elements gives the combined amplitude for Rydberg state $f$:
\begin{equation}
    \bra{\psi_f} T \ket{\Psi_o} = A_{f}+\bra{\psi_f} T^{(2)} \ket{\Psi_o}.
\end{equation}

The differential change of electron yield for autoionization of state $\ket{\psi_f}$ is obtained by taking the difference between the probability of a two photon transition and the initial probability in state $\ket{\psi_f}$. For initially populated states this is second order in the SWIR field strength:
\begin{equation}
    D(f,t_o) = -\mathcal{E}_o^2 2 \Re{A_{f}^{*} \bra{\psi_f} T^{(2)} \ket{\Psi_o} } +O(\mathcal{E}_o^4).
\end{equation}

Notice that even if the interaction with the intermediate states allows, in principle, for the excitation of $nf'$ states with different $J$ and $M$ values, as well as the excitation of the $np$ autoionizing states, their contributions are too weak to be observed in the experiment.
%This is supported by the fact that the frequencies involved in the experimental data involve only the energies of the $nf'$ states. 
Also, the differential signal is due to the interference between the initially populated states \emph{only}. The depletion and enhancement is thus polarized in the direction of pump pulse, irrespective of the Raman probe polarization.

\begin{figure}
\includegraphics[width=\linewidth]{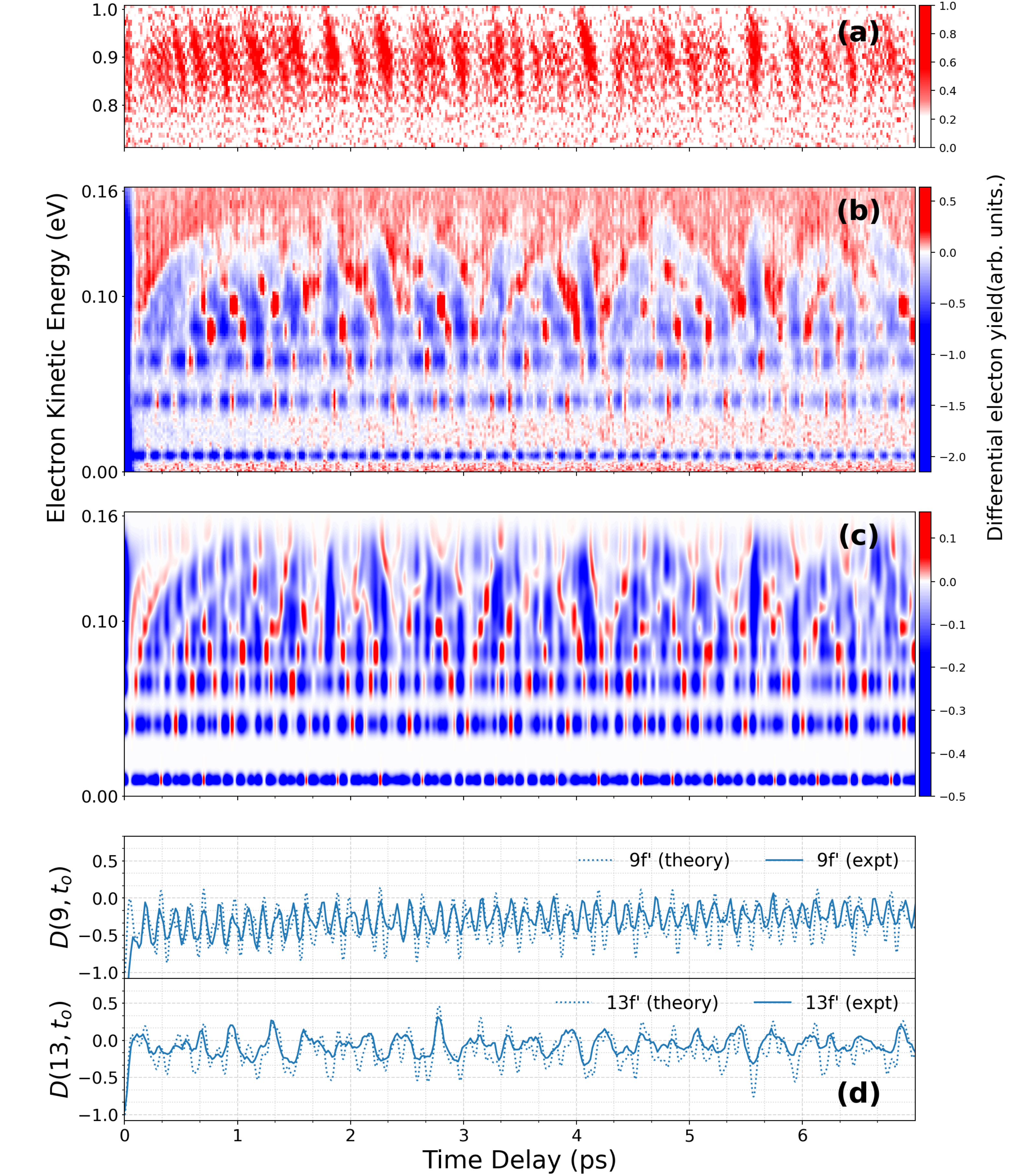}
\caption{Experimental differential electron spectrograms corresponding to (a) photoionization and (b) autoionization. (c) Theoretical results for the autoionization. The photoionization exhibits beats due to the interferences between different $nf'$ ionization paths, but there is loss of resolution along the kinetic energy axis due to the probe bandwidth. The autoionization yield is modulated by the coherent Raman redistribution induced by the probe. The detection following long autoionization timescales provides excellent resolution in electronic binding energies, allowing much better analysis of the beats and wavepacket evolution. (d) Lineouts from (b) and (c) to quantitatively compare experimental and theoretical beat evolution at $9f'$ and $13f'$ states.}
\label{fig:Timedelay}
\end{figure}

Fig. ~\ref{fig:Timedelay}(a) shows the traditional photoelectron spectrogram resulting from ionization of $nf'$ wavepacket with the femtosecond SWIR probe, which interrupts the autoionization. The electron kinetic energy is along vertical axis and pump-probe time-delay along the horizontal axis.  For the 1~eV SWIR probe, the electron kinetic energy lies between 0.8-1~eV. Due to the broad bandwidth of the probe pulse, the photoionization amplitudes from different states composing the wavepacket can interfere in the continuum and lead to beating as a function of pump-probe delay. The beat frequencies depend on the energy separation between constituent states, but there is no state-resolved information along the kinetic energy axis. 
%In order for the electron amplitudes from two (or more) bound states to interfere, the ionizing probe pulse must have enough spectral bandwidth to overlap the photoelectron lines originating from each state. 
The spectral smearing, and therefore loss of kinetic energy resolution, is a necessary feature of such experiments since, without it, there would be no interference. 

In contrast, Fig. ~\ref{fig:Timedelay}(b) shows the experimental results obtained using our Raman interferometry approach, where we monitor the differential changes in the autoionization yield due to the coherent redistribution of the wavepacket amplitudes by the probe pulse. For the $nf'$ states, autoionized electrons emerge with between 0 and 180 meV of kinetic energy, and these states have $\approx 10-1000$ ps lifetimes. The deferred detection inherent in this method yields impressive kinetic energy resolution, and we can monitor the quantum beating at specific electronic binding energies of the system. For example, the lowest kinetic energy feature, corresponding to $9f'$ state, shows a faster oscillation due to larger spacing between the lower $n$ Rydberg levels.

Another important feature of this experiment is that it demonstrates coherent control of the wavepacket composition via redistribution of electronic probability by the Raman pulse. Note that the blue color in Fig. ~\ref{fig:Timedelay} implies reduction in the autoionization probability compared with initial value, whereas the red color implies higher autoionization yield at that specific energy. Thus, we observe that as the time delay is varied the amplitudes of certain states are enhanced at the expense of others. 

Fig. ~\ref{fig:Timedelay}(c) shows the calculated differential electron yield spectrogram for parameters identical to the experiment. The agreement between experiment and theory is remarkable, showing that the model captures the detailed evolution of the wavepacket dynamics as well as the light-induced enhancement of specific wavepacket components. Fig. ~\ref{fig:Timedelay}(d) shows the lineouts of differential electron yield at kinetic energies corresponding to the $9f'$ and $13f'$  states, to quantify the experiment-theory agreement.  

\begin{figure}
\centering
\includegraphics[width=8cm]{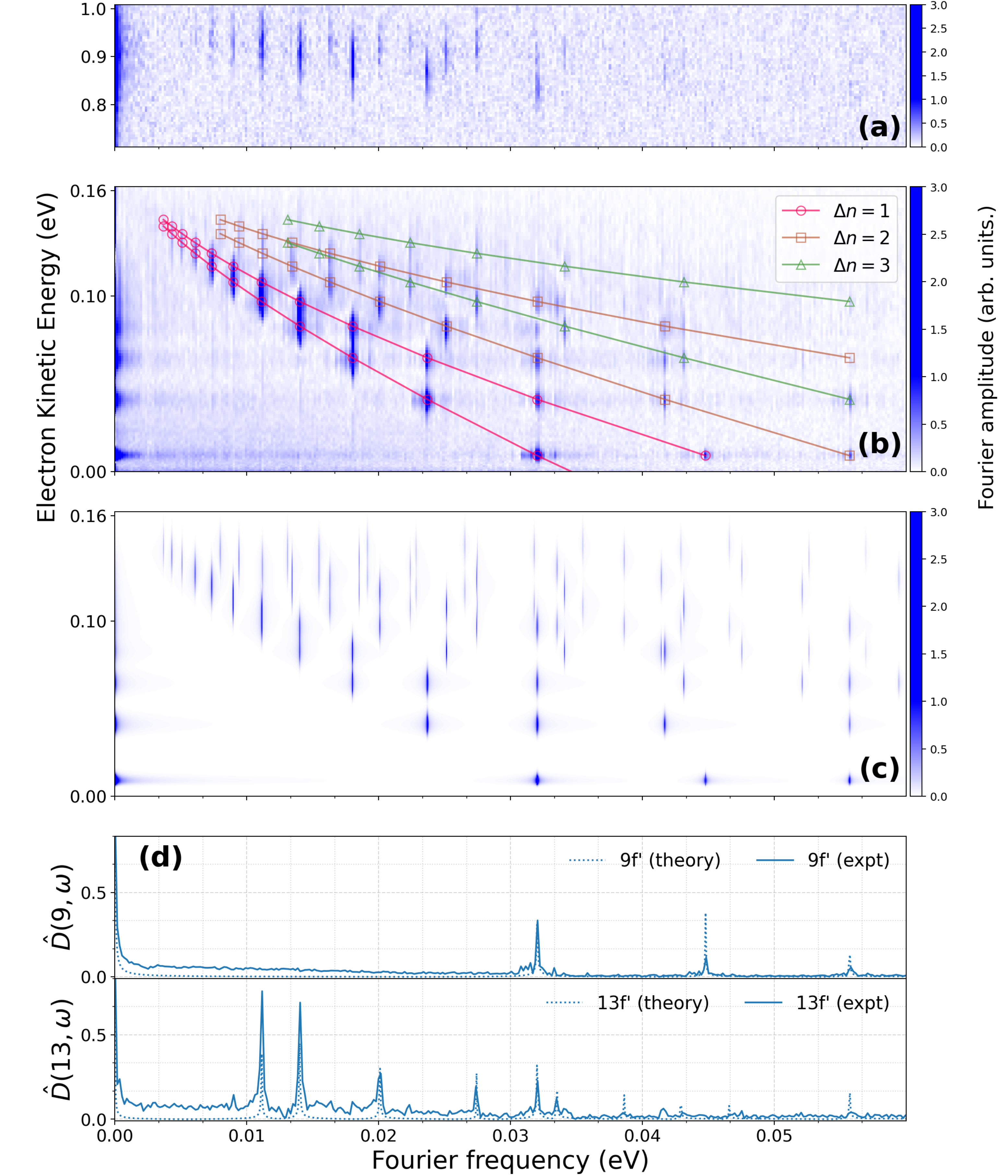}
\caption{Fourier spectrogram of the (a) photoionization experiment, (b) autoionization experiment, and (c) autoionization theory. The Fourier frequencies corresponding to the energy differences between $nf'$ states appear on the horizontal axis, while  peaks corresponding the electron binding energies are visible along the vertical axis. Photoionization results do not identify specific states unambigiously, whereas the differential autoionization signals from the Raman probing allow the investigation of electronic composition in detail. Overlaid curves represent the expected position of the beats for pairs of states separated by $\Delta n=\pm1,2,3$. (d) Lineouts from (b) and (c) to quantitatively compare experimental and theoretical beat strengths at $9f'$ and $13f'$ energies. }
\label{fig:FFTspecs}
\end{figure}

Fig.~\ref{fig:FFTspecs} shows the Fourier transform of the time-evolution reported in Fig.~\ref{fig:Timedelay}.  Fig.~\ref{fig:FFTspecs}(a) again refers to direct interferences in the higher energy continuum due to photoionization by the femtosecond probe pulse. One can see quantum beats due to energy separations between the states. However, due to kinetic energy smearing by the broadband probe, one cannot determine the binding energies or separate nearby beats. In contrast, the differential autoionization signal in Fig.~\ref{fig:FFTspecs}(b) exhibits superb resolution along the kinetic energy axis limited only by the instrumental resolution. We can clearly identify the electronic states along the vertical axis ordered by the quantum number from $9f'$ onwards. It is also easy to observe that several beat frequencies on the horizontal axis correspond to more than one pair of states interfering with each other. The curves overlaid on Fig.~\ref{fig:FFTspecs}(b) identify the beats corresponding to principle quantum number differences $\Delta n=\pm1,2,3$. The high frequency beats, absent in direct photoionization, are prominently visible here due to better sensitivity. The theoretical result in  Fig.~\ref{fig:FFTspecs}(c) captures the experimental detail in terms of both the beat frequencies and kinetic energies. This agreement allows us to quantify the electronic properties of autoionizing states and demonstrates the benefits of our approach over direct photoelectron interferometry. 

\begin{figure}
\includegraphics[width=8cm]{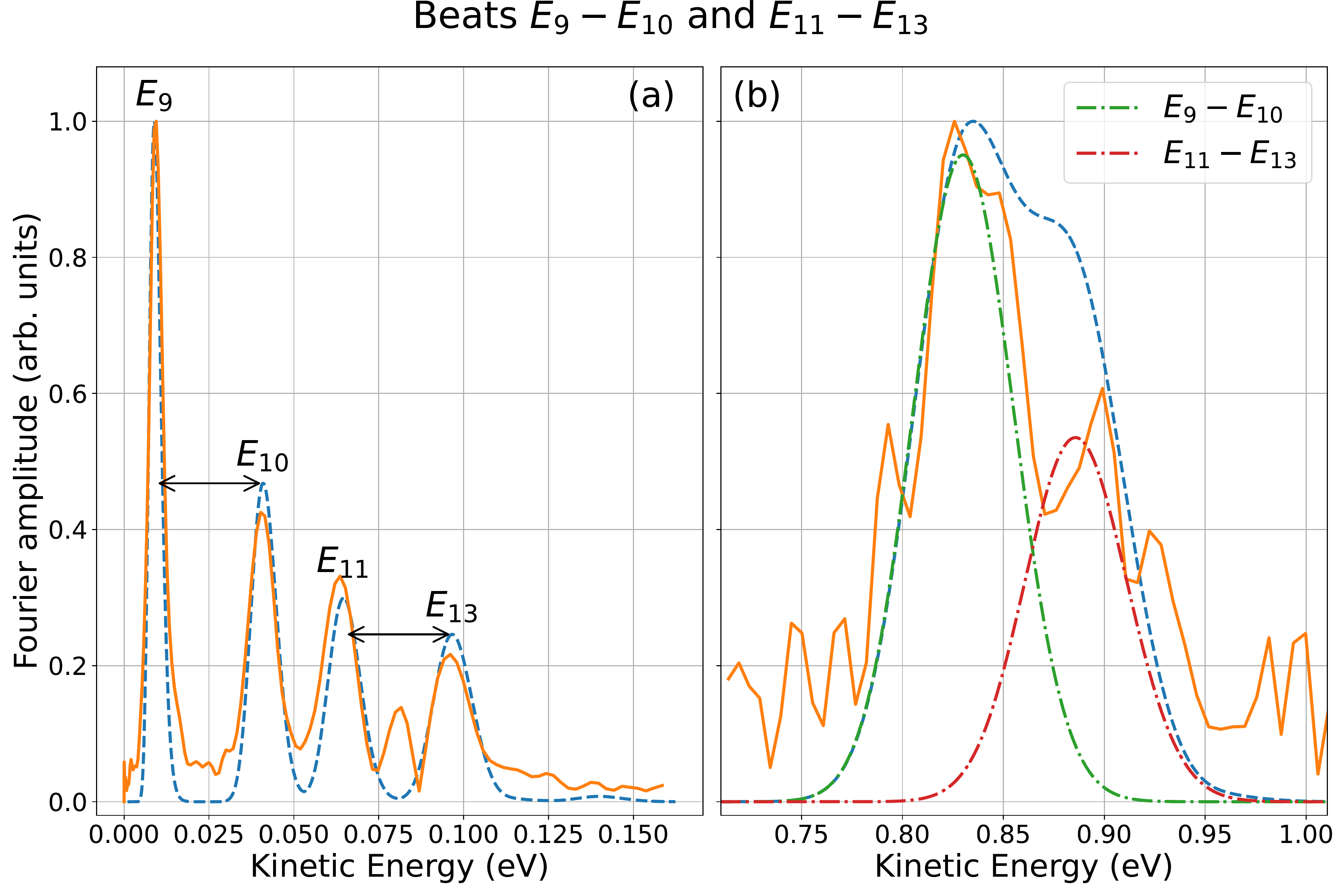}
\caption{(a) Kinetic energy resolved amplitude of the differential autoionization signal in experiment (solid line) and theory (dashed line) differentiates two beat pairs, between states $9f'$ and $10f'$ and states $11f'$ and $13f'$, that lie in the same beat frequency bin and otherwise would be non-separable. (b) Photoionization spectra do not have the kinetic energy resolution to distinguish the overlapping contributions in experiment (solid line) or theory (dashed line). Calculated photoionization amplitude for each beat pair is shown  (dot-dashed line).}
\label{fig:Resolve}
\end{figure}

Fig.~\ref{fig:Resolve} provides specific examples of the advantage of our method. Fig.~\ref{fig:Resolve}(a) shows kinetic energy resolved Fourier amplitudes extracted from Fig.~\ref{fig:FFTspecs} at the beat frequency $0.032$ eV. The plot clearly identifies the participating electronic states and demonstrates that our narrow frequency bin actually contains two pairs of beats, $E_{9}- E_{10}$ and $E_{11}-E_{13}$, separated only by $\sim$2$\mu$eV, which otherwise would be indistinguishable due to the limitation on the Fourier frequency resolution. Note that the small peak at 0.08 eV corresponds to the state $12f'$, since beat $E_{12}-E_{15}$ also falls in the same frequency bin for the experimental data. Fig.~\ref{fig:Resolve}(b) shows the kinetic energy resolved Fourier amplitude at the same beat frequency, but in the photoionization energy region. Clearly, one cannot distinguish specific contributions from the experimental data due to the lack of kinetic energy resolution and poor signal-to-noise. Thus, the assignment of states in this case is impractical with conventional photoelectron interferometry.
%The theoretical results also exhibit a single feature from which identification of composing states can be non-unique. 
Even the theoretical plot for the photoionization cannot unambiguously resolve the two beat contributions. Due to finite bandwidth of the laser, a single broad peak results from the overlap of the photoionization amplitudes of two pairs of states. The amplitude for each beat pair is centered around the average energy for the pair, as shown by the dot-dashed lines.

This experiment showcases the ability to conduct a femtosecond-resolved study with a very high energy resolution, where we clearly distinguish the electronic states that otherwise would have produced overlapping beats in the Fourier spectrum of a traditional photoelectron signal. The essence of our technique lies in utilization of ultrashort probe pulse to drive Raman interferences between the states, and by separately conducting the detection on long timescales by monitoring the differential changes in the signal. By utilizing a tailored Raman probe, one can also exert coherent control to create the desired wavepackets. Detection via natural decay processes, such as autoionization, is not a strict requirement. One could alternatively employ a quasi-monochromatic pulse that photoionizes the wavepacket on long timescales, thus projecting the spectrally-resolved information into the continuum.  Our approach can be generalized to the excited molecular wavepackets that usually form a dense manifold of states, providing a powerful tool to analyze the time-evolution and electronic structure of complex systems.

%% Acknowledgments
%%
AP, JW, and AS acknowledge support from the National Science Foundation under award PHYS 1912455. MA and CHG acknowledge support from the U. S. Department of Energy, Office of Science, Office of Basic Energy Sciences, under Award No. DESC0010545

\bibliography{Ar_AI_refs}

\end{document}